\documentclass[aps,prl,twocolumn,groupedaddress]{revtex4}
\usepackage{graphicx}

\begin{document}


\title{Coherence measurements on Rydberg wave packets kicked by a half-cycle pulse}


\author{J. M. Murray, S. N. Pisharody, H. Wen, P. H. Bucksbaum}
\affiliation{FOCUS Center, Department of Physics, University of Michigan, Ann Arbor, MI
48109-1120}


\date{\today}

\begin{abstract}
A kick from a unipolar half-cycle pulse (HCP) can redistribute population and shift the relative
phase between states in a radial Rydberg wave packet. We have measured the quantum coherence
properties following the kick, and show that selected coherences can be destroyed by applying an
HCP at specific times. Quantum mechanical simulations show that this is due to redistribution of
the angular momentum in the presence of noise. These results have implications for the storage and
retrieval of quantum information in the wave packet.

\end{abstract}


\maketitle


Atoms can store and process data encoded in their quantum states
\cite{StokelyPRA02,StenholmPRA01,shapiroPRL03}. We previously investigated the storage of
information as quantum phase in Rydberg wave packets, and its retrieval using half-cycle pulses
(HCP's) \cite{RanganModOpt2002, rangan:033417, ahn, ahn2, ahn3}. The HCP produces multi-mode
interference between the states, converting phase information into state populations \cite{ahn3}.

More sophisticated quantum information processing involves
multiple operations on the same data register.  The phase information in the atom must then be retained after each
operation. Even though the HCP interaction is unitary,
technical noise problems that do not affect the populations may destroy phase information, thus
eliminating further processing capability. Loss of phase coherence in a Rydberg wave packet can be
caused by several factors, such as background electrical noise, atomic collisions, and radiative
decay. Here, we study the phase shifts and quantum coherence of a
Rydberg wave packet following a weak HCP.

We use a wave packet holography technique developed earlier for wave packet sculpting
\cite{weinachtNature99}, but not previously applied to HCP interactions. We excite a Rydberg wave
packet in atomic cesium followed by an HCP at a specific time delay. A subsequent laser pulse
excites a reference wave packet. Interference between the two wave packets is analyzed by
state-selective field ionization (SSFI), and the results show the correlation between the
populations in all pairs of states. We also view the same interference in the absence of the HCP,
and thereby determine the change induced in the relative phases of the states.

Cesium atoms from an effusive source are excited from the $6s$ ground state to the $7s$ launch
state by a two photon transition using 1079nm pulses from the focused output of a
Ti:sapphire-pumped optical parametric amplifier. A Rydberg wave packet is then excited from $7s$
to $n=28,\dots,32$, $\ell = 1$.  The excitation laser spectrum is shaped using an acousto-optic
Fourier-plane filter \cite{Warrenshaper}, so that the phases of the states are initially equal.
These phases evolve in time. After time $\tau$, a weak THz HCP is applied with the same
polarization as the two laser pulses. Details on the generation and detection of HCP's, as well as
their interaction with Rydberg states and wave packets have been reported previously
\cite{TielkingPRA95,noordamPRL97,dyou,jonesyou,craman,maedaPRL04}. The HCP used in our experiment
provides an impulse of 0.0014 a.u. (atomic units) to the Rydberg electron. The wave packet is then
probed by superposing a reference wave packet on the same atoms. The reference is identical to the
launch wave packet, but delayed up to 50 psec. In a typical run we alternate collecting coherence
data with the HCP on and off, to minimize the effects of laser drift or changes in the atomic
beam.

\begin{figure}
\includegraphics{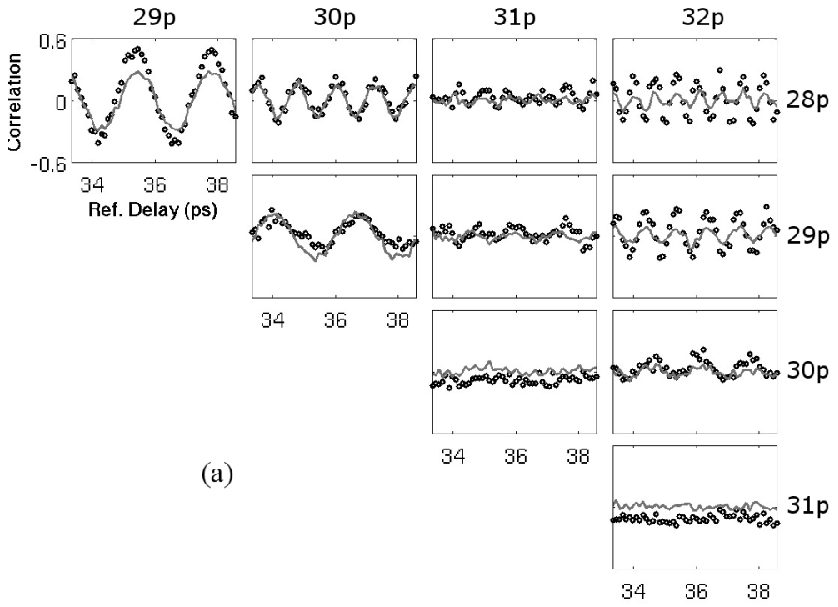}
\includegraphics{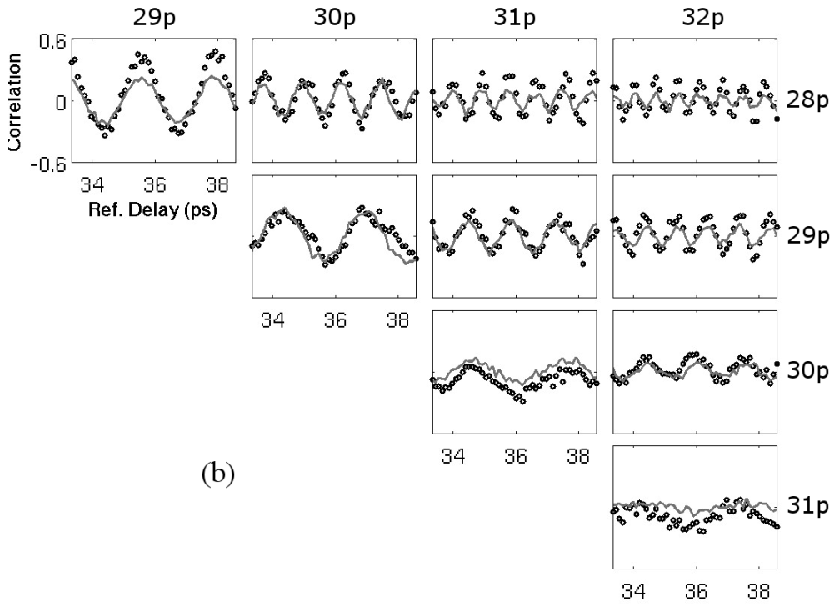}
\caption{\label{corr_comp} Selective destruction of coherence: (a)The result of a correlation
measurement with the HCP delayed 7.2 ps after the wave packet excitation. Note that the
correlation of the $31p$ state with all other states vanishes while the rest of the states have a
non-zero correlation with each other. The dark circles represent the experimental results while
the thick gray line is obtained from our simulations. (b) A similar plot of correlations but with
the HCP delayed 7.8 ps after wave packet excitation shows the $31p$ state having non-zero
correlation with the other states.}
\end{figure}

There is a remarkable loss of coherence between some pairs of states when kicked with an HCP at
specific times. These same states retain their coherence if they are kicked at later times (see
Fig. \ref{corr_comp}). We also observe a $\tau$-dependent shift in the phases of the correlations.
These results imply that if we use an HCP as an operator on a wave packet, the choice of delay
$\tau$ can make a significant difference in our ability to perform additional operations.

To gain a better understanding of the physical processes that lead to these phase-shifts and loss
of coherence, we constructed an impulse model of an HCP that simulates the experimental
conditions. The states $\Psi_k(\vec{r},t)=\Psi_k(\vec{r})e^{-i\omega_kt}$ of the Rydberg wave
packet have different energies $\hbar \omega_k$, so that the wave packet has a time dependence:
\begin{equation}\label{wavepacket}
\Psi(\vec{r},t)=\sum_k C_k \Psi_k(\vec{r}) e^{-i (\omega_k t-\phi_k)}.
\end{equation}
Here $C_k$ and $\phi_k$ are the amplitude and phase, respectively, corresponding to $
\Psi_k(\vec{r},t) $.

The superposition of two arbitrary wave packets excited from the same launch state and separated in
time by a delay $\tau$ can be expressed at time $t$ after the arrival of the first wave packet as
\begin{equation}\label{wpinterference}
\sum_k \Psi_k(\vec{r}) e^{-i \omega_k t}\left(C_{k1}e^{i \phi_{k1}}+C_{k2}e^{-i
((\omega_g-\omega_k)\tau-\phi_{k2})}\right)
\end{equation}
where $\omega_g$ is the energy of the launch state. The populations in the states of the combined
wave packet vary with the delay $\tau$ as \small
\begin{eqnarray}\label{poptau}
    P_k(\tau) = C_{k1}^2+C_{k2}^2+2C_{k1}C_{k2}\cos(\phi_{k1}-\phi_{k2}-(\omega_k-\omega_g)\tau)
    \nonumber \\
    =(C_{k1}^2+C_{k2}^2)\left[1+C_k'\cos(\phi_{k1}-\phi_{k2}-(\omega_k-\omega_g)\tau)\right]
\end{eqnarray}
\normalsize where $C_k'=2C_{k1}C_{k2}/(C_{k1}^2+C_{k2}^2)$.

The relative phase between pairs of states is measured using a covariance technique
\cite{weinachtNature99}. The correlation between the populations in various states in the wave
packet is represented by
\begin{equation}\label{correlationexpr}
    r_{jk}(\tau) = \left[\frac{\overline{P_j.
    P_k}-\overline{P_j}.\overline{P_k}}{\sigma_j\sigma_k}\right]_\tau
\end{equation}
where
\begin{equation}
 \sigma_j\sigma_k = \sqrt{\overline{P^2_j}-\overline{P_j}^2}.\sqrt{\overline{P^2_k}-\overline{P_k}^2}
 \end{equation}

Using Eqn(\ref{poptau}) and averaging over multiple optical cycles, we can expand the terms of the
numerator in Eqn(\ref{correlationexpr}):
\begin{widetext}
\begin{eqnarray}\label{corr-numexp}
\overline{P_j}.\overline{P_k} &=&(C_{j1}^2+C_{j2}^2)(C_{k1}^2+C_{k2}^2) \nonumber \\
\overline{P_j.P_k}&=&(C_{j1}^2+C_{j2}^2)(C_{k1}^2+C_{k2}^2)\left[1+\frac{C_j'C_k'}{2}
\cos((\phi_{j1}-\phi_{k1})-(\phi_{j2}-\phi_{k2})-(\omega_j-\omega_k)\tau)\right] \nonumber \\
\Rightarrow \overline{P_j.P_k}-\overline{P_j}.\overline{P_k}
&=&2C_{j1}C_{j2}C_{k1}C_{k2}\cos((\phi_{j1}-\phi_{k1})-(\phi_{j2}-\phi_{k2})-(\omega_j-\omega_k)\tau)
\end{eqnarray}
\end{widetext}

The denominator in Eqn(\ref{correlationexpr}) can be expressed as
\begin{equation}
 \sigma_j\sigma_k = 2C_{j1}C_{j2}C_{k1}C_{k2}.
\end{equation}

We can then write the expected correlation between the states of the wave packet,
\begin{equation}\label{correlation}
    r_{jk}(\tau)=\cos((\phi_{j1}-\phi_{k1})-(\phi_{j2}-\phi_{k2})-(\omega_j-\omega_k)\tau).
\end{equation}

The presence of any background noise in the measurements modifies the correlations as
\begin{equation}\label{correlationmeas}
    r_{jk}(\tau)_{meas}=\sqrt{\left(1-\frac{\sigma_{N_j}^2}{\sigma_{j_{meas}}^2}\right)
    \left(1-\frac{\sigma_{N_k}^2}{\sigma_{k_{meas}}^2}\right)}
    r_{jk}(\tau)
\end{equation}
where $\sigma_{N_j}$ represents the standard deviation of the noise present in the measurement of
state $j$ and $\sigma_{j_{meas}} =\sqrt{\sigma_j^2+\sigma_{N_j}^2}$ represents the measured
standard deviation for population identified as state $j$.

\begin{figure}
\includegraphics{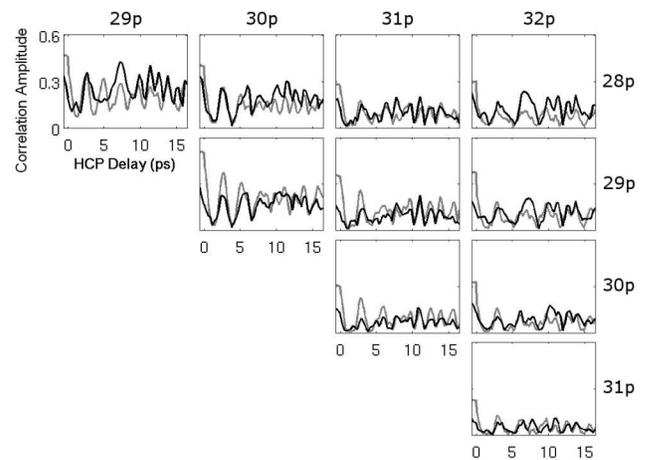}
\caption{\label{corramps_comp}Comparison of correlation amplitudes: The amplitude of the
correlation measurements from the experiment are compared with the results of our simulations for
a range of delays of the HCP following wave packet excitation. The dark line represents the
experimental results while the thick gray line is a result of the simulations.}
\end{figure}

\begin{figure}
\includegraphics{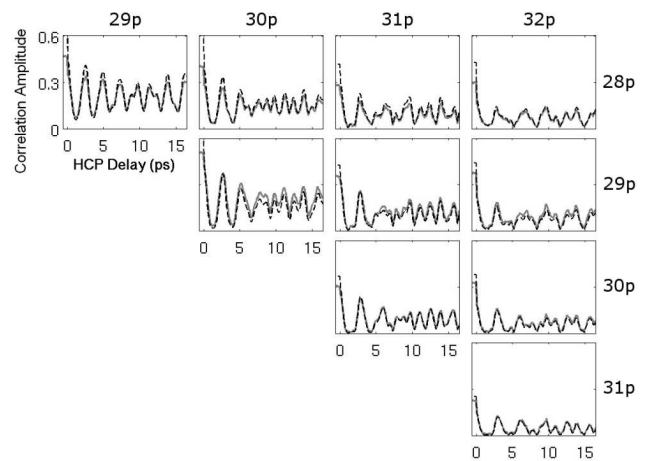}
\caption{\label{pstatepops} $p$-state populations vs. correlation amplitude: The product of the
$p$-state amplitudes (dashed line) is plotted along with the correlation amplitude (solid gray
line) for each pair of $p$-states.}
\end{figure}

The simulations are in good agreement with our experimental results (see Fig. \ref{corr_comp}).
The observed variation in the amplitude of the correlations is reproduced well by the simulation
and is shown in Fig. \ref{corramps_comp}. Our simulations suggest that the loss of coherence in
the experiment is tied to the redistribution of $p$-states into other angular momentum states by
the HCP. Fig. \ref{pstatepops} shows the remaining $p$-state population following an HCP at
various delays after the wave packet excitation and compares the correlation amplitude to the
product of $p$-state amplitudes. The similarity in the two curves confirms that the correlation
amplitude is determined largely by the population remaining in the $p$ states. The peaks in the
correlation amplitudes correspond to times when the electron is close to the core, while the dips
occur when the electron is far from the core. This is consistent with the classical notion that
the maximal angular momentum transfer occurs when the impulse is exerted far from the core.

The loss of coherence when angular momentum is transferred out of the p-states is almost entirely
attributable to the background noise that is always present in the measurement. In the absence of
any noise, the correlation amplitudes are always unity. The effect of noise is introduced in the
simulation as random fluctuations in the signal level detected for each state. The correlation
amplitudes in the simulation match the experimental results well when we introduce shot-to-shot
fluctuations of 100$\%$ rms. This suggests a remarkable robustness of the correlation method to
retrieve useful information in the presence of large signal fluctuations.

\begin{figure}
\includegraphics[width=3.3in]{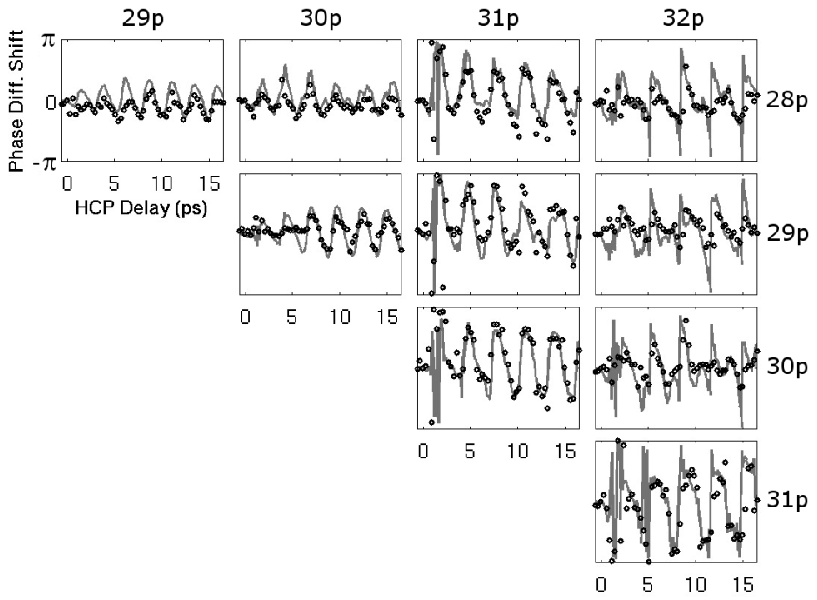}
\caption{\label{phasecomp} Phase change comparison : The change in relative phase of the
correlation as a function of HCP delay as measured in experiment (black circles) is compared with
the phase of the correlations obtained from our simulation (solid gray line).}
\end{figure}

The HCP-induced state redistribution changes the relative phases in the wave packet. The measured
phase changes are plotted in Fig. \ref{phasecomp} together with the corresponding results from
simulations. All the major features in the data are reproduced by the simulations. The agreement
is poorest when the amplitude of the correlations is small, at times when the wave packet is near
the outer turning point of its orbit.

These correlation measurements show that a Rydberg wave packet can maintain its coherence
following excitation by an HCP. The good agreement with simulations shows that the HCP can induce
controllable phase shifts between components of the wave packet, which can be measured
subsequently using interference with a reference wave packet. Finally, we have found that the HCP
can selectively alter the angular momentum of individual states in the wave packet, while
preserving the angular momentum and maintaining overall coherence among the rest of the states.
The modulations in the amplitude and phase of the measured correlations as a function of the delay
of the HCP can be used to characterize the effect of any HCP operator acting on phase information
stored in the Rydberg atom data register.

Previous work by Ahn et. al \cite{ahn,ahn2,ahn3} has demonstrated the possibility of information
storage in the phase relationship between states in a Rydberg wave packet and its retrieval by an
HCP. The current work demonstrates that we can use an HCP of programmable strength and delay as an
operator for manipulating the stored information.

The authors would like to thank C. Rangan and R. R. Jones for several fruitful discussions. This
work has been supported by the National Science Foundation under grant no.9987916 and the Army
Research Office under grant no. DAAD 19-00-1-0373.

\bibliography{corr-paperrefs}

\end{document}